\documentclass[aps,pra,twocolumn,showpacs,superscriptaddress]{revtex4-1}
\usepackage{amsmath,amssymb}
\usepackage[pdftex]{graphicx}
\usepackage{subfigure}
\usepackage{booktabs}
\usepackage{color}
\usepackage{braket}
\newcommand \beq{\begin{eqnarray}}
\newcommand \eeq{\end{eqnarray}}

\begin{document}

\title{Population-imbalance instability in a Bose-Hubbard ladder in the presence of a magnetic flux}
\author{Shun Uchino}
\affiliation{DQMP, University of Geneva, 24 Quai Ernest-Ansermet,
1211 Geneva, Switzerland}
\author{Akiyuki Tokuno}
\affiliation{ Coll\`ege de France, 11 place Marcelin Berthelot, 
F-75005 Paris, France}
\affiliation{Centre de Physique Th\'{e}orique, Ecole Polytechnique,
CNRS, F-91128 Palaiseau Cedex, France}
\date{\today}

\begin{abstract}
 We consider a two-leg Bose-Hubbard ladder in the presence of a magnetic
 flux. 
 We make use of Gross-Pitaevskii, Bogoliubov,  bosonization, and renormalization
 group approaches to reveal a structure of ground-state phase diagrams
 in a weak-coupling regime relevant to cold atom experiments.
 It is found that except for a certain flux $\phi=\pi$, the system
shows different properties as changing hoppings, which
also leads to a
 quantum phase transition similar to the ferromagnetic
 $XXZ$ model. This implies that population-imbalance instability occurs
for certain parameter regimes.
 On the other hand, for $\phi=\pi$, it is shown that an umklapp process
 caused by commensurability of a magnetic flux stabilizes a
 superfluid with chirality and the system does not experience such
 a phase transition.

\end{abstract}


\maketitle

\section{Introduction}
Quantum systems subject to high magnetic fields are known to 
acquire nontrivial characteristics such as
the Hofstadter butterfly~\cite{PhysRevB.14.2239} and
quantum Hall effect \cite{prange1987quantum} in two dimensional systems.
Recently, the so-called synthetic gauge
fields~\cite{RevModPhys.83.1523,goldman2014light,2014arXiv1410.8425G} 
available in ultracold atomic systems pave the way to realizations of
such systems.
An advantage of cold atoms is that one can control the geometry, dimension
or quantum statistics of systems and parameters of microscopic
Hamiltonians at unprecedented levels, which is used to explore
non-trivial quantum states.
Along these lines of researches, the Hofstadter
Hamiltonian~\cite{PhysRevLett.111.185301,PhysRevLett.111.185302} 
and Haldane topological model~\cite{jotzu2014experimental} have been
realized in cold atoms.

In addition to such a non-trivial feature of non-interacting quantum
matters, as is well known, an interaction is a key ingredient for the
diversity of nature.
Indeed various phenomena such as superconductivity, superfluidity and
Mott transition are understood as a consequence of the interactions. 
One can thus expect the interaction effects in such topological matters
to cause further non-trivial nature, and effective approaches
incorporating interactions are required in theory.

In one dimension one can successfully apply field theoretical approaches
to incorporate interaction effects in a non-perturbative manner.
Thus, reduction of dimensions would be a way to understand physics
involving magnetic fields and correlations. 
The minimal model to show non-trivial effects in the presence of
magnetic fields is a two-leg ladder, and the bosonic version was
first discussed in Ref.~\cite{PhysRevB.64.144515} in the context of
Josephson junction arrays.
In this study, it has been predicted that two different phases show up:
Meissner and vortex phases.
While in the former phase a chiral current analogous to a Meissner edge
current is induced on the legs by a magnetic flux, in the latter it is
significantly reduced due to penetration of vortices, which is analogous
to field-induced vortices in type-II superconductors.
Later on, the theoretical interest has been devoted to
the strong correlated regime, and it has been demonstrated that
commensurability of a particle filling poses a Mott insulator
with chirality and interesting critical
properties~\cite{PhysRevA.85.041602,PhysRevB.87.174501,PhysRevLett.111.150601,tokuno,PhysRevA.91.013629}.

Remarkably, the two-leg bosonic ladder subject to a magnetic flux has
been successfully realized in an experiment~\cite{atala2014observation},
where a weakly-interacting regime is concerned.
In this experiment, it has been confirmed that behaviors of chiral
currents are consistent with what has been predicted in
Ref.~\cite{PhysRevB.64.144515}.
Thus it seems that a basic consensus in a
weak-coupling regime is obtained.

More recently, however, it has been argued in
Ref.~\cite{PhysRevA.89.063617} that in a weak-coupling regime there
should exist an additional phase where a spontaneous population
imbalance between the legs occurs.
This additional phase named a biased ladder phase has been shown with
 the theory of weakly-interacting Bose gases normally used in higher
dimensions~\cite{pethick2002bose}.
At the same time, in one dimension quantum fluctuations should be non-negligible
in most cases.
Thus it is worth considering whether the biased ladder phase is still
robust against quantum fluctuations. 

\begin{figure}[t]
 \begin{center}
  \includegraphics[width=1\linewidth]{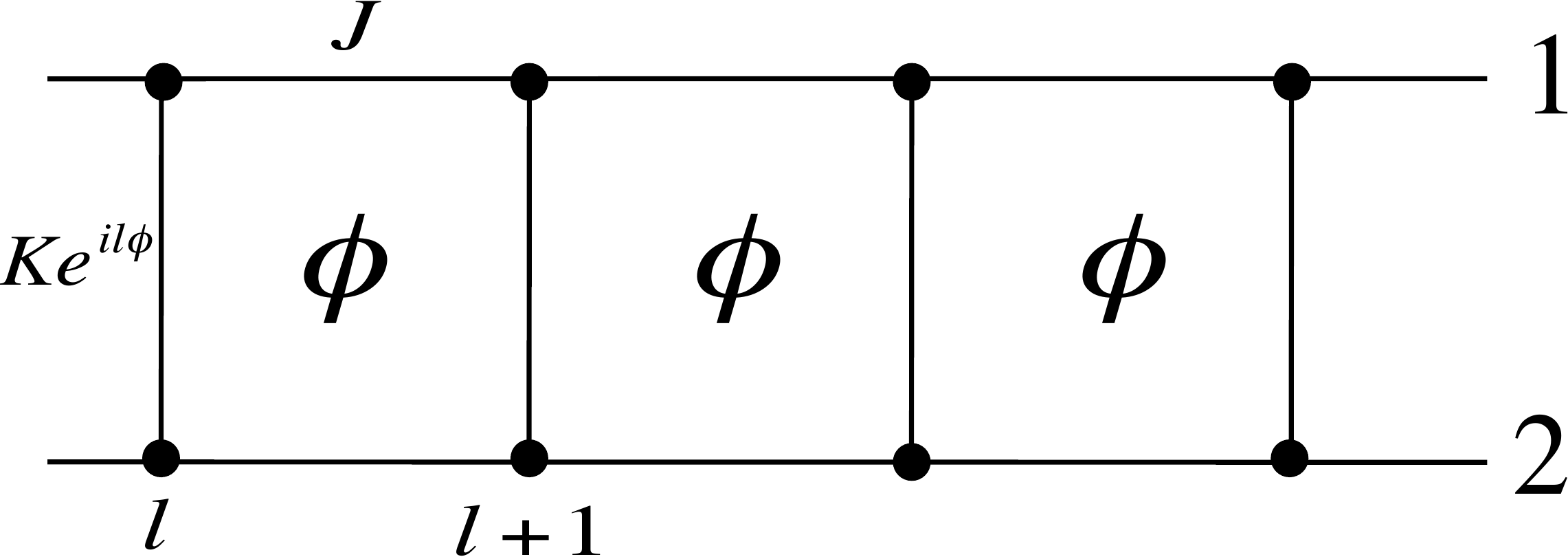}
  \caption{A schematic figure of a two-leg Bose-Hubbard model with a
  flux $\phi$.
  In the gauge chosen here, the flux effect appears only in rung
  hoppings.}
  \label{fig1}
 \end{center}
\end{figure}

In this paper, we examine the two-leg Bose-Hubbard model in the presence
of a magnetic flux in a weakly-interacting regime by means of a couple
of effective-theory approaches. 
We show that a spontaneous population imbalance indeed occurs, and is
stable against quantum fluctuation effects.
In particular, we point out that the effective theory has the similarity
to a ferromagnetic $XXZ$ quantum spin model. 
This implies that the Heisenberg point exists in the phase diagram,
where $SU(2)$ symmetry shows up in the low-energy effective theory
although the original Hamiltonian does not possess that symmetry.
This situation is somewhat similar to a two-leg extended Bose-Hubbard
system analyzed in Ref~\cite{PhysRevA.81.053606}, where the
low-energy effective theory possesses an emergent symmetry.

We also state that umklapp processes coming from commensurability of a flux
should be seriously considered
at the mean-field level, which has been overlooked in the previous
studies.
The umklapp processes existing in $\phi=\pi$ destabilize the biased
ladder phase, and as a consequence, only the commensurate vortex phase is
allowed. 

The structure of the paper is as follows.
In Sec.~\ref{sec:formulation}, we review structures of single-particle
bands as a function of $\phi$, and low-energy effective Hamiltonian
reflecting the band structure.
The single-particle band bottom shows different topologies: A single
minimum for a small flux, and double minima for a large flux.
In Secs.~\ref{sec:single-minimum} and~\ref{sec:double-minimum} we discuss
physics separately for a single-minimum and for a double-minimum
band structure, in which Meissner, vortex, and biased ladder phases are
allowed depending on $\phi$ and $K/J$. 
In Sec.~\ref{sec:summary} summary and
perspective on a  phase transition
between the Meissner and biased ladder phases, and on a stronger interaction effect are provided.
Technical details on renormalization group equations are
addressed in the Appendix.

\section{Formulation of the Problem}
\label{sec:formulation}
Following the setup in Ref.~\cite{atala2014observation}, we consider the
following two-leg Bose-Hubbard ladder Hamiltonian: 
\beq
H &&=-J\sum_{l=1}^L\sum_{p=1,2}(e^{iA^{\parallel}_{l,p}}b^{\dagger}_{l+1,p}b_{l,p}+\mathrm{H.c.})
\nonumber\\
  &&\quad -K\sum_{l}(e^{iA^{\perp}_{l}}b^{\dagger}_{l,1}b_{l,2}+\mathrm{H.c.})
+\frac{U}{2}\sum_{l,p}n_{l,p}(n_{l,p}-1),
\label{eq:hamiltonian}
\eeq
where $J$ and $K$ are the hopping amplitudes along the leg and rung
directions, respectively.
The applied flux is introduced via the Peierls substitution, and the
corresponding gauge fields along the chain and rung directions are
denoted by $A^{\parallel}_{l,p}$ and $A^{\perp}_{l}$, respectively.
The flux $\phi$ is then given as
$\phi=A^{\parallel}_{l,1}-A^{\perp}_{l+1}-A^{\parallel}_{l,2}+A^{\perp}_{l}$.
The technology of laser-assisted
tunneling~\cite{PhysRevLett.107.255301,PhysRevLett.108.225303,PhysRevLett.108.225304,PhysRevLett.111.185301,PhysRevLett.111.185302}
generates spatially-dependent phase in rung hoppings, which leads to the
$\phi$ flux per plaquette as described in Fig.~\ref{fig1}.
Namely, in this paper we choose the following gauge as
$A^{\parallel}_{l,p}=0$ and $A^{\perp}_{l}=\phi l$.
Taking into account the fact that an interatomic interaction is given by
an $s$-wave scattering length, and the stability in bosonic systems, we
restrict ourselves to a local repulsive interaction, $U>0$. 

Apparently, the Hamiltonian~\eqref{eq:hamiltonian} is invariant under
the simultaneous transformations, $b_{l,1(2)}\to b_{l,2(1)}$ and
$\phi\to-\phi$.
Thus, we can safely take the domain of definition in $\phi$
as $0<\phi\le\pi$.

By using the general relation between current and Hamiltonian,
$j=-\frac{\partial H}{\partial A}$ where $A$ is the gauge field,
we can define current operators along the legs and rungs as
\beq
&&j^{\parallel}_{l,p}=iJ(b^{\dagger}_{l+1,p}b_{l,p}
-b^{\dagger}_{l,p}b_{l+1,p}),\label{eq:leg-current}\\
&&j^{\perp}_{l}=iK\left(e^{il\phi}b^{\dagger}_{l,1}b_{l,2}-
e^{-il\phi}b^{\dagger}_{l,2}b_{l,1}\right). \label{eq:rung-current}
\eeq
For the sake of convenience, we also introduce chiral current
along the legs:
\beq
j_c=j^{\parallel}_{l,1}-j^{\parallel}_{l,2}. \label{eq-chiral-current}
\eeq
We will see that $j_c$ and $j^{\perp}$ play important roles in
characterizing each phase. 

Since we are interested in the regime $J,K\gg U$, we start with
diagonalizing the single-particle Hamiltonian.
To this end, we perform gauge and Fourier transformations as 
$b_{l,1}=\frac{1}{\sqrt{L}}\sum_k e^{i(k+\frac{\phi}{2})l}b_{k,1}$
and $b_{l,2}=\frac{1}{\sqrt{L}}\sum_k e^{i(k-\frac{\phi}{2})l}b_{k,2}$.
Then, by considering a unitary transformation for $b_{k,1}$ and
$b_{k,2}$ 
\beq
\begin{pmatrix}
b_{k,1}\\
b_{k,2}
\end{pmatrix}
=
\begin{pmatrix}
\cos\left(\frac{\xi_k}{2}\right) & -\sin\left(\frac{\xi_k}{2}\right)\\
\sin\left(\frac{\xi_k}{2}\right) & \cos\left(\frac{\xi_k}{2}\right)
\end{pmatrix}
\begin{pmatrix}
\alpha_k\\
\beta_k
\end{pmatrix},
\eeq
where
\beq
\sin\left(\frac{\xi_k}{2}\right)
=-\sqrt{\frac{1}{2}\left(1-
\frac{\sin\left(\frac{\phi}{2}\right)\sin k}{
\sqrt{\left(\frac{K}{2J}\right)^2+\sin^2\left(\frac{\phi}{2}\right)\sin^2k}}
\right)},\nonumber\\ \\
\cos\left(\frac{\xi_k}{2}\right)
=\sqrt{\frac{1}{2}\left(1+
\frac{\sin\left(\frac{\phi}{2}\right)\sin k}{
\sqrt{\left(\frac{K}{2J}\right)^2+\sin^2\left(\frac{\phi}{2}\right)\sin^2k}}
\right)},\nonumber\\
\eeq
the single-particle Hamiltonian can be
diagonalized~\cite{PhysRevB.73.195114,PhysRevB.76.195105} as
\beq
H_0=\sum_k(E_+(k)\alpha^{\dagger}_k\alpha_k+E_{-}(k)\beta^{\dagger}_k
\beta_k).
\eeq
Here, the single-particle spectrum is given by
\beq
E_{\pm}(k)=2J\left[-\cos\left(\frac{\phi}{2}\right)\cos k
\pm\sqrt{\left(\frac{K}{2J}\right)^2
+\sin^2\left(\frac{\phi}{2}\right)\sin^2k}\right],
\label{eq:band}
\nonumber\\
\eeq
which depicts the two-band structure as well as the $2\pi$ periodicity,
reflecting the two-leg ladder geometry.

Unless a strong interaction is concerned, single-particle low-energy
states, bottoms in the lowest band, play important roles in the
low-energy many-body states.
With this understanding, we neglect effects of the higher band
$\alpha_k$ by keeping in mind the condition $J,K\gg U$.

\begin{figure}[h]
 \begin{center}
  \includegraphics[width=1\linewidth]{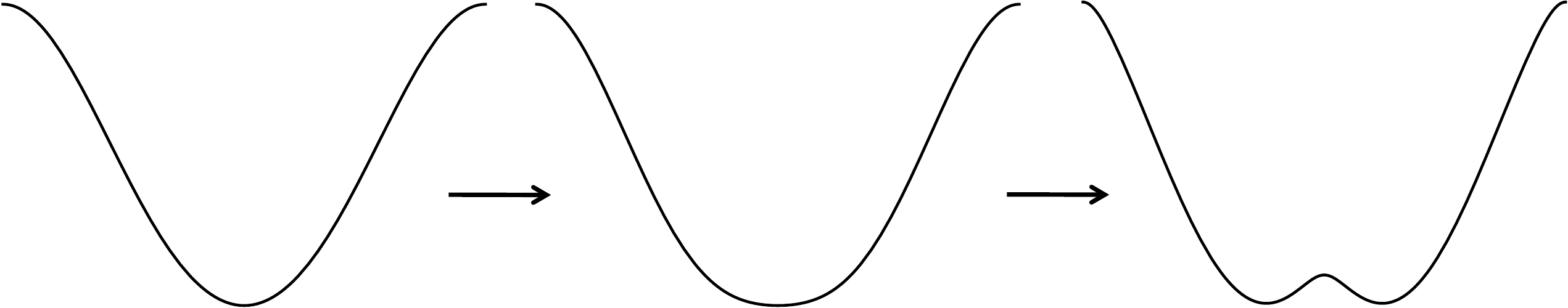}
  \caption{Changes of topology in the lower band at a certain
  $\phi\ne\pi$.
  The direction of the arrows means the reduction of $K/J$.
  In a strong enough $K/J$, the band has a single minimum
  while in the opposite limit, the band of double well structure forms.
  In between, there exists a critical point in which the band bottom
  becomes quartic in $k$. 
  The double-well structure is always maintained at $\phi=\pi$
  regardless of values of $K/J$.}
  \label{fig2}
 \end{center}
\end{figure}
Let us look into a behavior of the lower band in more detail.
We first obtain extrema via
$\frac{\partial E_-(k)}{\partial k}=0$,
which leads to
\beq
\sin
k\left[\cos\left(\frac{\phi}{2}\right)-\frac{\sin^2\left(\frac{\phi}{2}
\right)\cos k}
{\sqrt{\left(\frac{K}{2J}\right)^2+\sin^2\left(\frac{\phi}{2}\right)\sin^2k}}
\right]=0.\nonumber\\
\label{eq:minim}
\eeq
If $\phi\ne\pi$ with $K/J\gg1$, $k=0$ and $\pm \pi$ are the solution of
Eq.~\eqref{eq:minim}, and  $k=0$ gives the minimum of the band.
As in the case of the normal cosine band, the dispersion near $k=0$ is
approximated to be quadratic in $k$. 
It is straightforwardly shown that the condition for the single-minimum
structure can be expressed as~\cite{PhysRevB.73.195114,PhysRevB.76.195105}
\beq
 \left(\frac{K}{2J}\right)^2
 >\frac{\sin^4\left(\frac{\phi}{2}\right)}{1-\sin^2\left(\frac{\phi}{2}\right)}.
 \label{eq:single-min}
\eeq
As $K/J$ decreases, on the other hand, the double-well structure starts
to show up.
The critical point between the single- and double-minimum structure is
given by
\beq
 \left(\frac{K}{2J}\right)^2
 =\frac{\sin^4\left(\frac{\phi}{2}\right)}{1-\sin^2\left(\frac{\phi}{2}\right)},
 \label{eq:critical_K/J}
\eeq
which means the equality limit of the inequality in
Eq.~\eqref{eq:single-min}.
Then the dispersion near the bottom becomes quartic in $k$.
In the double-well structure case, two separated minima $\pm Q$ can be
obtained from the factor of the square bracket in
Eq.~\eqref{eq:minim}~\cite{PhysRevB.73.195114,PhysRevB.76.195105}, 
\beq
Q=\sin^{-1}\left[\sqrt{\sin^2\left(\frac{\phi}{2}\right)-\left(\frac{K}{
2J}\right)^2\cot^2\left(\frac{\phi}{2}\right)}\right].
\eeq
In addition, the dispersions around the minimum point are quadratic in
$k$ as in the case of the single minimum.
We emphasize that $\phi=\pi$ is special because the symmetric
double-well structure is strongly protected and its minima are 
located at $Q=\pm\frac{\pi}{2}$ regardless of $K/J$.
% These can be easily checked from the facts that
% the right hand side of Eq. \eqref{eq:single-min}
% diverges and thus it is impossible to satisfy the inequality,
% and $\cot\left(\frac{\pi}{2}\right)=0$.

The similar behavior is also found in the case of changes of $\phi$ with
a fixed $K/J$. 
In that case it is shown that for the small enough $\phi$, the band
forms single-minimum structure, and shows up the double-well structure
when $\phi$ goes through a critical value $\phi_c$.
A critical flux $\phi_c$ between these two structures is shown to
be~\cite{PhysRevB.73.195114,PhysRevB.76.195105}
\beq
\sin^{-1}\left(\frac{\phi_c}{2}\right)
=\sqrt{\frac{\sqrt{\left(\frac{K}{J}\right)^4
+16\left(\frac{K}{J}\right)^2}
-\left(\frac{K}{J}\right)^2}{8}}.
\eeq

Based on the change of the band structure discussed above, 
let us incorporate interaction effects within a weakly-interacting
regime, $J,K\gg U$.
In this interaction regime, the bottoms of the band are also important
for bosonic systems, which is an essentially different point from the
fermion systems.
Therefore, we first truncate all the effects involving the upper band
($\alpha_k$). 
The Hamiltonian~(\ref{eq:hamiltonian}) is reduced to
\beq
H=\sum_kE_-\beta^{\dagger}_k\beta_k+\frac{1}{2L}\sum_{k_1,k_2,k_3,k_4}
\Gamma_{k_1,k_2,k_3,k_4}\beta^{\dagger}_{k_1}\beta^{\dagger}_{k_2}
\beta_{k_3}\beta_{k_4},
\nonumber\\
\label{eq:original-h}
\eeq
where
\beq
\Gamma_{k_1,k_2,k_3,k_4}
&=&U\sum_{n'\in\mathbb{Z}}\delta_{k_1+k_2-k_3-k_4,2\pi n'}
\nonumber \\
&&
\times \Big[\sin\left(\frac{\xi_{k_1}}{2}\right)
\sin\left(\frac{\xi_{k_2}}{2}\right)
\sin\left(\frac{\xi_{k_3}}{2}\right)
\sin\left(\frac{\xi_{k_4}}{2}\right)
\nonumber\\
&&
+\cos\left(\frac{\xi_{k_1}}{2}\right)\cos\left(\frac{\xi_{k_2}}{2}\right)
\cos\left(\frac{\xi_{k_3}}{2}\right)
\cos\left(\frac{\xi_{k_4}}{2}\right)
\Big].\nonumber\\
\label{eq:gamma}
\eeq
Note that in contrast with a system in continuum space, we need to consider
scattering processes involving a finite momentum transfer equal to the
reciprocal lattice vector. 
In our model, the finite momentum transfer to $2\pi n'$ with 
an integer $n'$ is allowed, which is nothing but the umklapp process,
and turns out to play a crucial role in the $\phi=\pi$ case.

In what follows, we separately look into many-body ground states in each 
topology of the single-particle band.

\section{Band with a single minimum}
\label{sec:single-minimum}
For weakly-interacting bosons in higher dimensions, the Gross-Pitaevskii
(GP) approach as one of the mean-field theories is known to provide good
results~\cite{pethick2002bose}. 
As a consequence, the system undergoes a Bose-Einstein condensate (BEC),
which also implies spontaneous breaking of $U(1)$ symmetry.
Thereby, a gapless excitation mode known as a Nambu-Goldstone (NG) mode
is obtained. 

However, it is also well known that for an interacting one-dimensional
bosonic system, there exists neither BEC nor NG mode in the
thermodynamic limit. 
Namely the mean-field analysis underestimates fluctuation effects, and
cannot thus capture correct results in the one-dimensional cases.
However, one-dimensional superfluids in the weakly-interacting regime
show very slow power-law decay in such a way that the
order effect works almost comparably to the quantum fluctuation.
In addition, there is also an acoustic phonon mode similar to the NG
mode, although it does not correspond to the spontaneous symmetry breaking. 
From these facts, it is found that the GP approach is not correct in a strict sense, 
but would provide a practically reasonable
starting point  to discuss the ground state and low-energy excitation structure.
In addition the advantage of the GP approach is that both 
kinetic and interaction energies can be
simultaneously taken into account at the mean-field level~\footnote{In
the single minimum case, however, the kinetic energy is vanished due to
the occupation at $k=0$. In Sec. \ref{sec:double-minimum}, we see an
example where the kinetic energy takes a nonzero contribution.}.

Based on the above observations, let us consider the system with the single band.~\cite{tokuno}
As far as the weakly-interacting bosons are concerned, the bosons dominantly populate
the minimum of the lower energy band, and one can perform an approximation such that
all the energy states except for ones in the vicinity of the minimum are projected out.
Thus the low-energy single-particle spectrum is approximated as
\beq
E_{-}(k)\approx -E_0+\frac{k^2}{2M},
\eeq
where $E_0=K+2J\cos\left(\frac{\phi}{2}\right)$ and
$\frac{1}{M}=\frac{d^2E_-(k=0)}{dk^2}$.
Since all the wave numbers are restricted to be $|k_j|\ll 1$ due to the
long-wave-length approximation, the effective interaction parameter
$\Gamma_{k_1,k_2,k_3,k_4}$ in Eq.~\eqref{eq:gamma} is approximated in
the following way: The small wave number $k$ leads to
$\sin(\xi_{k}/2)\approx -1/\sqrt{2}$ and
$\cos(\xi_{k}/2)\approx 1/\sqrt{2}$, and by substituting the
approximated form of $\xi_{k}$ into Eq.~\eqref{eq:gamma}, the
interaction parameter is approximated as $\Gamma_{k_1,k_2,k_3,k_4}\approx U/2$.
Thus, we reach the following low-energy effective Hamiltonian~(\ref{eq:original-h}):  
\beq
H\approx\int dx\Big[-\beta^{\dagger}(x)\frac{\nabla^2}{2M}\beta(x)
+\frac{U}{4}
\beta^{\dagger}(x)\beta(x)
\beta^{\dagger}(x)\beta(x)\Big],\nonumber\\
\label{eq:continuum-h}
\eeq
where 
$\beta(x)=\frac{1}{\sqrt{L}}\sum_k\beta_ke^{ikx}$.
We note that this is essentially identical to the Lieb-Liniger
model~\cite{PhysRev.130.1605}. 
As far as the weak-coupling limit is concerned, one may consider  
the following GP ground state:
\beq
|GS\rangle=\frac{1}{\sqrt{N!}}(\beta^{\dagger}_{k=0})^N|0\rangle,
\eeq
where $N$ is the number of bosons.

Let us next incorporate long-wavelength fluctuations
which play crucial roles in low-energy properties.
To this end, we adopt the hydrodynamic approach also known as
the bosonization for
bosons~\cite{haldane1981luttinger,giamarchi2003quantum,cazalilla2004bosonizing}:
\beq
\beta(x)\sim\Big[n-\frac{\nabla\varphi(x)}{\pi}\Big]^{\frac{1}{2}}\sum_{m\in\mathbb{Z}}
e^{2im[\pi nx-\varphi(x)]}e^{-i\theta(x)},
\label{eq:bosonized-form}
\eeq
where $n=N/L$ is the mean density.
We introduced the density and phase fluctuations, $\varphi(x)$ and
$\theta(x)$, respectively, and the commutation relation between them is
given by 
\beq
[\theta(x),\frac{1}{\pi}\nabla\varphi(x')]=i\delta(x-x').
\eeq
Applying the above bosonization formula~\eqref{eq:bosonized-form} to
Eq.~\eqref{eq:continuum-h}, we obtain
\beq
H_{\text{eff}}=\frac{v_0}{2\pi}\int dx
 \left[\frac{1}{K_0}(\nabla\varphi)^2+K_0(\nabla\theta)^2\right],
\eeq
where $v_0=\sqrt{\frac{nU}{2M}}$ and $K_0=\pi\sqrt{\frac{2n}{MU}}$.
This is the Hamiltonian for the celebrated Tomonaga-Luttinger liquid
(TLL), which corresponds to a $c=1$ conformal field theory. 
It is remarkable that in this TLL Hamiltonian, the long-range order
(LRO) of the single-particle density matrix, incorrectly predicted by the
GP mean-field theory, is directly confirmed to be modified into a correct quasi-LRO of the
algebraic decay~\cite{giamarchi2003quantum}; 
$\langle\beta^{\dagger}(x)\beta(0)\rangle\sim\left(\frac{1}{x}\right)^{1/2K_0}$.

Let us next look into the rung~\eqref{eq:rung-current} and chiral
current~\eqref{eq-chiral-current} by translating them
in effective theory derived above.
By using the bosonization formula~\eqref{eq:bosonized-form} the current
operators, Eqs.~\eqref{eq:rung-current} and~\eqref{eq-chiral-current},
are expressed as 
\beq
&&j^{\perp}(x)\sim0, \\
&&j_{c}(x)\sim 2nJ\sin\left(\frac{\phi}{2}\right)+O(\nabla^2\theta),
\label{eq:chiral-m}
\eeq
where we point out that the rung current vanishes regardless of bosonization,
while the chiral current has the nonzero constant term and the terms
starting from $\nabla^2\theta$.
Note that at the level of the long-wave approximation,
a fluctuation of the rung current $j^{\perp}$ disappears.
On the other hand, a fluctuation of the chiral current,
$\delta{j_c}\equiv j_c-2nJ\sin\left(\phi/2\right)\sim \nabla^2 \theta$, behaves as
\beq
 \langle{\delta{j_c}(x)\delta{j_c}(0)}\rangle
 \sim 1/x^4,
 \eeq
 which is given by the Gaussian property in the TLL Hamiltonian such
 that 
$\langle\nabla^2\theta(x)\nabla^2\theta(0)\rangle \sim 1/x^4$.
% This implies that a \textit{current fluctuation}, $\delta{j}_c=$
% \beq
% \langle \delta j(x)\delta j(0)\rangle 
% =\langle j(x)j(0)\rangle-\langle j(x) \rangle
% \langle j(0)\rangle ,
% \eeq 
% is going to be zero in the rung current.
% In the chiral current, it 
% decays as $1/x^4$, which originates from the Gaussian properties in the TLL
% Hamiltonian such that
% $\langle\nabla^2\theta\rangle=0$ and
% $\langle\nabla^2\theta(x)\nabla^2\theta(0)\rangle \sim 1/x^4$.
In addition, Eq.~\eqref{eq:chiral-m} shows that the chiral current
increases with $\phi$. 
These properties correspond to the Meissner phase introduced in
Ref.~\cite{PhysRevB.64.144515}, which has been derived in a
condition $J\gg K,U$ different from the present case.

\section{Band with double minima}
\label{sec:double-minimum}
We next examine low-energy properties of the system with the double-well
band structure where we can distinguish a commensurate wave number $Q$,
giving the lowest-energy single-particle states, from incommensurate one. 
In the commensurate cases, $Q$ can be represented as 
$Q=\pi p/q$, where $p,q$ are coprime numbers.

The effect of commensurability is related to types of
interactions.
As pointed out in Ref.~\cite{PhysRevB.64.144515}, a $q$-body interaction produces the
serious effect for $Q=\pi p/q$.
Thus, if arbitrary multi-body interactions come into the low-energy
effective theory, every commensurability should be taken care.
Note that it does not mean that multi-body interactions are required at
the microscopic level.
Namely even if only two-body interactions are assumed in the
microscopic Hamiltonian, multi-body interactions are generated as
virtual multiple-scattering processes when we integrate
out irrelevant high-energy degrees of freedom such as deriving a
low-energy effective Hamiltonian and implementing a perturbative
renormalization group theory.

However, such virtual multiple-scattering processes would be suppressed
in the weakly-interacting case, and the relevant case would be only for
$Q=\pi/2$ in which the two-body interaction yields the strong
commensurability effect.

Here we first consider an incommensurate $Q$ case in Sec.~\ref{sec:IC}.
Next in Sec.~\ref{sec:Q=pi/2}, we move on to the discussion of the
$Q=\frac{\pi}{2}$ ($\phi=\pi$) case as one of the commensurate cases.
The other commensurability is also briefly discussed in
Sec.~\ref{sec:otherC}. 

\subsection{Incommensurate $Q$ case}\label{sec:IC}
In contrast with the single minimum case, the mean ground-state density
with the double well structure depends on the couplings of the
Hamiltonian.
Following the analysis for a BEC on a double-well
potential~\cite{pethick2002bose}, 
we assume the following ansatz, first introduced
in Ref.~\cite{PhysRevA.89.063617}:
\beq
|GS\rangle=\frac{1}{\sqrt{N!}}(e^{i\theta_+}
\cos\gamma\beta^{\dagger}_{Q}+e^{i\theta_-}\sin\gamma
\beta^{\dagger}_{-Q})^N|0\rangle, 
\label{eq:gp}
\eeq
where $\gamma$ and $\theta_{\pm}$ are variational parameters.

By taking the expectation value of $H$ in Eq.~\eqref{eq:original-h} with
the above ansatz, one obtains~\cite{PhysRevA.89.063617} 
\beq
&&\frac{E_0(\gamma,\theta_{\pm})}{N}=E_-(Q)+\frac{Un}{4}\Big[
\left(\frac{3}{2}\sin^2\xi_{Q}-1\right)
\sin^22\gamma\nonumber\\
&& \ \ \ \  \ \ \ \ \ \ \ \ \ \ \ \ -\sin^2\xi_{Q}+2\Big], \label{eq:mf-energy}\\
&&\langle n_{+}\rangle=\langle \beta^{\dagger}_+(x)\beta_+(x)\rangle
=n\cos^2\gamma,\\
&&\langle n_{-}\rangle=\langle \beta^{\dagger}_{-}(x)\beta_{-}(x)\rangle
=n\sin^2\gamma,
\eeq
where $\beta_{\pm}(x)=\frac{1}{\sqrt{L}}\sum_k \beta_{\pm Q+k}e^{ikx}$.
We note that these mean-field values have no dependence in
$\theta_{\pm}$, which implies that the problem is reduced to
optimization of the single variational parameter, $\gamma$.
Then, the optimized $\gamma$ is alternatively determined
by whether~\cite{PhysRevA.89.063617} 
\beq
\frac{3}{2}\sin^2\xi_{Q}<1, \label{eq:case1}
\eeq
or
\beq
\frac{3}{2}\sin^2\xi_{Q}>1. \label{eq:case2}
\eeq

In the former case~\eqref{eq:case1}, the ground state is minimized
by $\gamma=\pi/4$~\cite{PhysRevA.89.063617},
where the populations at $k=\pm Q$ are the same:
$\langle n_+\rangle=\langle n_-\rangle$. 
Thus, at the mean-field level, we expect that there are two independent
BECs in the ground state. 
We note that this is different from a BEC on a double-well potential,
where the ground-state energy depends on the relative phase
via a hopping term between the condensates~\cite{pethick2002bose}.
By contrast, such a hopping does not exist in our system, and thus 
the mean-field ground state is free to the relative phase.
We will come back to this point in an analysis in the $\phi=\pi$ case,
where a relative phase dependence shows up via the umklapp process
in a nontrivial manner.

In the latter case~\eqref{eq:case2}, the ground state is characterized
by $\gamma=0$ or $\pi/2$~\cite{PhysRevA.89.063617},
where the mean density becomes 
$(\langle n_+\rangle,\ \langle n_-\rangle)=(n,0)$ or $(0,n)$.
This is the solution such that all the bosons occupy either at $k=Q$ or
at $k=-Q$.
Thus, the mean-field theory shows that $Z_2$ symmetry is spontaneously
broken in the ground state, and a single BEC occurs simultaneously. 

The transition between these mean-field ground states occurs
at
$\frac{3}{2}\sin^2\xi_{Q}=1$,
which turns out to be rewritten as
\beq
\left(\frac{K}{2J}\right)^2=
\frac{\sin^4(\phi/2)}{\frac{3}{2}-\sin^2(\phi/2)}.
\label{eq:boundary-v-z2}
\eeq
What is important is that the above critical $K/J$ is
smaller than another critical $K/J$, given in
Eq.~\eqref{eq:critical_K/J}, between the single- and double-minimum band
topology.
Namely it means that the solution~\eqref{eq:boundary-v-z2} always exists
in the regime of $K/2J$ and $\phi$ in which the double-minimum band
structure comes out.
Therefore, we see that the transition between the mean-field ground
states always occurs at a certain $K/2J$ given by $\phi>0$.

Let us next look into fluctuation effects based on the above mean-field 
analyses. 
As in the case of the single-minimum band,
we approximate the Hamiltonian as~\cite{tokuno}
\beq
&&H\approx
\int dx\Big[-
\sum_{j=\pm }\beta^{\dagger}_{j}(x)\frac{\nabla^2}{2M^*}\beta_{j}(x)
\nonumber\\
&&+\frac{U(2+\sin^2\xi_{Q})}{8}(n_++n_-)^2
+\frac{U(2-3\sin^2\xi_Q)}{8}(n_+-n_-)^2
\Big],\nonumber\\
\label{eq:hamiltonian-dw}
\eeq
where $\frac{1}{M^*}=\frac{d^2E_-(\pm Q)}{dk^2}$ is the effective mass.

We first incorporate fluctuation effects in the case of the mean density
$\langle n_+\rangle=\langle n_+\rangle=n/2$ which is stable when
Eq.~\eqref{eq:case1} is obeyed.
By using bosonization formula~\eqref{eq:bosonized-form}
in $\beta_{\pm}$,
\beq
\beta_{\pm}(x)\sim\Big[\frac{n}{2}-\frac{\nabla\varphi_{\pm}(x)}{\pi}\Big]^{\frac{1}{2}}\sum_{m\in\mathbb{Z}}
e^{2im[\pi nx/2-\varphi_{\pm}(x)]}e^{-i\theta_{\pm}(x)},\nonumber\\
\eeq
with the density and phase fluctuation fields in the vicinity of
the bottoms $k=\pm Q$,
$\varphi_{\pm}$ and $\theta_{\pm}$,
we obtain
\beq
H_{\text{eff}}=\sum_{\nu=s,a}\frac{v_i}{2\pi}\int dx
\Big[K_{\mu}(\nabla\theta_{\mu})^2+\frac{(\nabla\varphi_{\mu})^2}{K_{\mu}}
\Big]\nonumber\\
+\frac{2g}{(2\pi\alpha)^2}\int dx \cos(\sqrt{8}\varphi_a),
\label{eq:bosonized-vx}
\eeq
where we have introduced the symmetric (anti-symmetric) fields,
$\varphi_{s(a)}=\frac{1}{\sqrt{2}}(\varphi_{+}+(-)\varphi_-)$ and
$\theta_{s(a)}=\frac{1}{\sqrt{2}}(\theta_{+}+(-)\theta_-)$.
For convenience's sake, we have also introduced the cutoff parameter $\alpha=1/(\pi n)$
\cite{giamarchi2003quantum}.
The velocities, $v_{s}$ and $v_{a}$, and TLL parameters, $K_{s}$ and
$K_{a}$, appearing in the quadratic parts of the Hamiltonian are
specified by 
\beq
&&v_s=\sqrt{\frac{nU(2+\sin^2\xi_Q)}{4M^*}}, \label{eq:vs}\\
&&v_a=\sqrt{\frac{nU(2-3\sin^2\xi_Q)}{4M^*}}, \label{eq:va}\\
&&K_s=\sqrt{\frac{n}{M^*U(2+\sin^2\xi_Q)}}, \label{eq:Ks}\\
&&K_a=\sqrt{\frac{n}{M^*U(2-3\sin^2\xi_Q)}}, \label{eq:Ka}
\eeq
and the coupling of the cosine term is given by
\beq
g=U\sin^2\xi_Q.
\eeq

In the effective Hamiltonian~\eqref{eq:bosonized-vx}, the symmetric and
anti-symmetric fields are decoupled.
Especially, the symmetric part is the conventional TLL Hamiltonian.
On the other hand, the anti-symmetric part seems not to be the TLL
Hamiltonian due to the presence of the cosine term.
Thus, the low-energy properties of the system are determined by the
relevancy of this cosine term in the sense of the renormalization group.
To this end, we implement the perturbative renormalization group analysis.
By treating the coupling constant $g$ as a perturbative parameter,
we obtain (See Appendix.), 
\beq
\frac{d(g/v_a)}{dl}=2(1-K_a)g/v_a,
\label{eq:rg-invortex}
\eeq
where $l$ is the scaling parameter.
In general, $K_s,K_a\gg1$ in weakly-coupling bosons with a contact
interaction.
This means that $g$ is the irrelevant coupling, and the cosine
term goes away in the low-energy limit.
Therefore, this phase is found to be characterized by the
two-independent TLL.

Let us next look into the currents.
The bosonized expressions of the operators are summarized as
\begin{widetext}
\beq
j_c(x)&\sim& nJ\Big[4\sin^2\left(\frac{\xi_Q}{2}\right)\sin
\left(\frac{\phi}{2}+Q\right)
-\sqrt{2}\sin^2\left(\frac{\xi_Q}{2}\right)\cos
\left(\frac{\phi}{2}+Q\right)\nabla\theta_a\nonumber\\
&&+\sqrt{2}\cos^2\left(\frac{\xi_Q}{2}\right)\cos
\left(\frac{\phi}{2}-Q\right)\nabla\theta_a
-4\sin\left(\frac{\xi_Q}{2}\right)\cos\left(\frac{\xi_Q}{2}\right)
\sin\left(\frac{\phi}{2}\right)\cos\left(Q(2x+1)-\sqrt{2}\theta_a\right)
\Big],\\
j^{\perp}(x)&\sim&nK\left(\sin^2\left(\frac{\xi_Q}{2}\right)
-\cos^2\left(\frac{\xi_Q}{2}\right)\right)\sin(2Qx-\sqrt{2}\theta_a).
\eeq
\end{widetext}
By taking the averages of the these quantities,
we obtain 
\beq
\langle j_c(x)\rangle&\sim&4nJ\sin^2\left(\frac{\xi_Q}{2}\right)\sin
\left(\frac{\phi}{2}+Q\right)
\label{eq:jc_incommQ} \\
\langle j^{\perp}(x)\rangle&\sim&0,
\label{eq:jperp_incommQ}
\eeq
where we used
$\langle\nabla\theta_a\rangle=\langle\cos(Q(2x+1)-\sqrt{2}\theta_a)\rangle
=\langle\sin(2Qx-\sqrt{2}\theta_a)=0$.
The form of Eqs.~\eqref{eq:jc_incommQ} and~\eqref{eq:jperp_incommQ} is
the same as what Wei and Mueller \cite{PhysRevA.89.063617} have derived
within the mean-field analysis for the net chiral current~\footnote{We
note an important difference between the mean-field and bosonization
approaches. In the mean-field approach, since $\theta_a$ (and $\theta_s$)
is ordered, the
local currents oscillate in space. In the bosonization approach, since
the anti-symmetric sector is  described by the TLL, the local currents
do not show such an oscillation as far as the commensurate effect does
not show up.}.
As shown in Refs.~\cite{PhysRevB.64.144515,PhysRevA.89.063617}, the
chiral current monotonically decreases with $\phi$, and goes to zero as
approaching $\phi\to\pi$.
This phase corresponds to the (incommensurate) vortex phase first
introduced in Ref.~\cite{PhysRevB.64.144515} in which the reduction of
the chiral current has been attributed to the penetration of the
vortices.
Indeed, a signature of the vortices is found in the correlations of the
current fluctuations,
$\delta j_c\equiv j_c-\langle j_c\rangle$ and
$\delta j^{\perp}\equiv  j^{\perp}-\langle j^{\perp}\rangle$.
They are calculated as 
\begin{widetext}
\beq
\langle\delta j_c(x)\delta j_c(0)\rangle&\sim&
\frac{n^2J^2}{K_a}
\Big[\sin^2\left(\frac{\xi_Q}{2}\right)\cos\left(\frac{\phi}{2}+
Q\right)+\cos^2\left(\frac{\xi_Q}{2}\right)\cos\left(\frac{\phi}{2}-
Q\right)\Big]^2\frac{1}{x^2}\nonumber\\
&&  
+8n^2J^2\sin^2\left(\frac{\xi_Q}{2}\right)
\cos^2\left(\frac{\xi_Q}{2}\right)\sin^2\left(\frac{\phi}{2}\right)
\cos(2Qx)\frac{1}{x^{1/K_a}}
\\
\langle\delta j^{\perp}(x)\delta j^{\perp}(0)\rangle&\sim&
n^2K^2\left[\sin^2\left(\frac{\xi_Q}{2}\right)
-\cos^2\left(\frac{\xi_Q}{2}\right)\right]^2
\cos(2Qx)\frac{1}{x^{1/K_a}},
\eeq
\end{widetext}
where we used the facts that
$\langle\nabla\theta(x)\nabla\theta(0)\rangle\sim\frac{1}{x^2}$,
$\langle e^{Ai\theta(x)}e^{-Ai\theta(0)} \rangle \sim\frac{1}{x^{\frac{A^2}{2K_a}}}$
with a constant $A$,
and
the cross terms such as
$\langle\nabla\theta(x)\cos(Q-\sqrt{2}\theta_a(0))\rangle$ vanish.
They have now oscillation components decaying with a power law.
Recalling $K_a\gg1$, these power-law decays are
extremely slow.

We next consider fluctuation effects in the case of the biased mean
density $(\langle n_+\rangle,\ \langle n_-\rangle)=(n,0)$ or $(0,n)$.
For the sake of simplicity, let us take the case of
$(\langle n_+\rangle,\ \langle n_-\rangle)=(n,0)$~\footnote{We can also
discuss the case of $(\braket{n_+},\braket{n_-})=(0,n)$ exactly in the
same manner, and the same result is obatined. However, only 
the magnetization has an opposite sign to Eq.~\eqref{eq:magnetization}}.
Namely all the bosons only populate around $k=Q$.
The degrees of freedom around $k=-Q$ are completely suppressed as far as
the interaction is weak enough, and  we may consider only the degrees of
freedom around $k=Q$.
Then the Hamiltonian simplified by the long-wave-length approximation is
given as
\beq
H=\int dx\Big[ -\beta^{\dagger}(x)\frac{\nabla^2}{2M^*}\beta(x)+
\frac{U(2-\sin^2\xi_Q)}{4}n^2\Big].
\eeq
Furthermore, by the bosonization formula~\eqref{eq:bosonized-form}, we
obtain
\beq
H_{\text{eff}}=\frac{\bar{v}}{2\pi}\int dx
\Big[\bar{K}(\nabla\theta)^2+\frac{(\nabla\varphi)^2}{\bar{K}}
\Big], 
\label{eq:Heff-BLP}
\eeq
where the velocity and TLL parameter are, respectively,
\beq
&&\bar{v}=\sqrt{\frac{nU(2-\sin^2\xi_Q)}{2M^*}},\\
&&\bar{K}=\sqrt{\frac{2n}{M^*U(2-\sin^2\xi_Q)}}.
\eeq
Unlike the equal density case, the effective theory is described by a
single TLL.

Let us next look at the currents.
They are bosonized as
given by
\begin{widetext}
\beq
j_c(x)&\sim&4nJ\Big[\sin^2\left(\frac{\xi_Q}{2}\right)
\sin\left(\frac{\phi}{2}+Q\right)
\Big]-2nJ\Big[\sin^2\left(\frac{\xi_Q}{2}\right)
\cos\left(\frac{\phi}{2}+Q\right)
-\cos^2\left(\frac{\xi_Q}{2}\right)
\cos\left(\frac{\phi}{2}-Q\right)
\Big]\nabla\theta,
\\
j^{\perp}(x)&\sim&0,
\eeq
where the rung current is shown to be zero at the level of
long-wave approximation as in the case of the Meissner phase.
The averages of them are calculated as 
\beq
\langle j_c(x)\rangle&\sim& 4nJ\sin^2\left(\frac{\xi_Q}{2}\right)
\sin\left(\frac{\phi}{2}+Q\right),\\
\langle j^{\perp}(x)\rangle&\sim&0,
\eeq
which are identical to the expressions for the net currents obtained in Ref.
\cite{PhysRevA.89.063617} and
look the same as those of the $\braket{n_+}=\braket{n_-}$ case,
i.e., Eqs.~\eqref{eq:jc_incommQ} and~\eqref{eq:jperp_incommQ}. 
However, this does not mean that all the low-energy properties coincide
with the equal-density case. 
Indeed, we find that the difference occurs in the current fluctuations
as
\beq
&&\langle\delta j_c(x)\delta j_c(0)\rangle\sim
\frac{2n^2J^2}{\bar{K}}\Big[\sin^2\left(\frac{\xi_Q}{2}\right)
\cos\left(\frac{\phi}{2}+Q\right)-\cos^2\left(\frac{\xi_Q}{2}\right)
\cos\left(\frac{\phi}{2}-Q\right)
\Big]^2\frac{1}{x^2} ,
\\
&&\langle\delta j^{\perp}(x)\delta j^{\perp}(0)\rangle\sim0.
\eeq
\end{widetext}
Thus, in contrast with the vortex phase in the $\langle
n_+\rangle=\langle n_-\rangle$ case, the current fluctuations in this
phase do not have an oscillating component.

\begin{figure}[t]
 \begin{center}
  \includegraphics[width=1\linewidth]{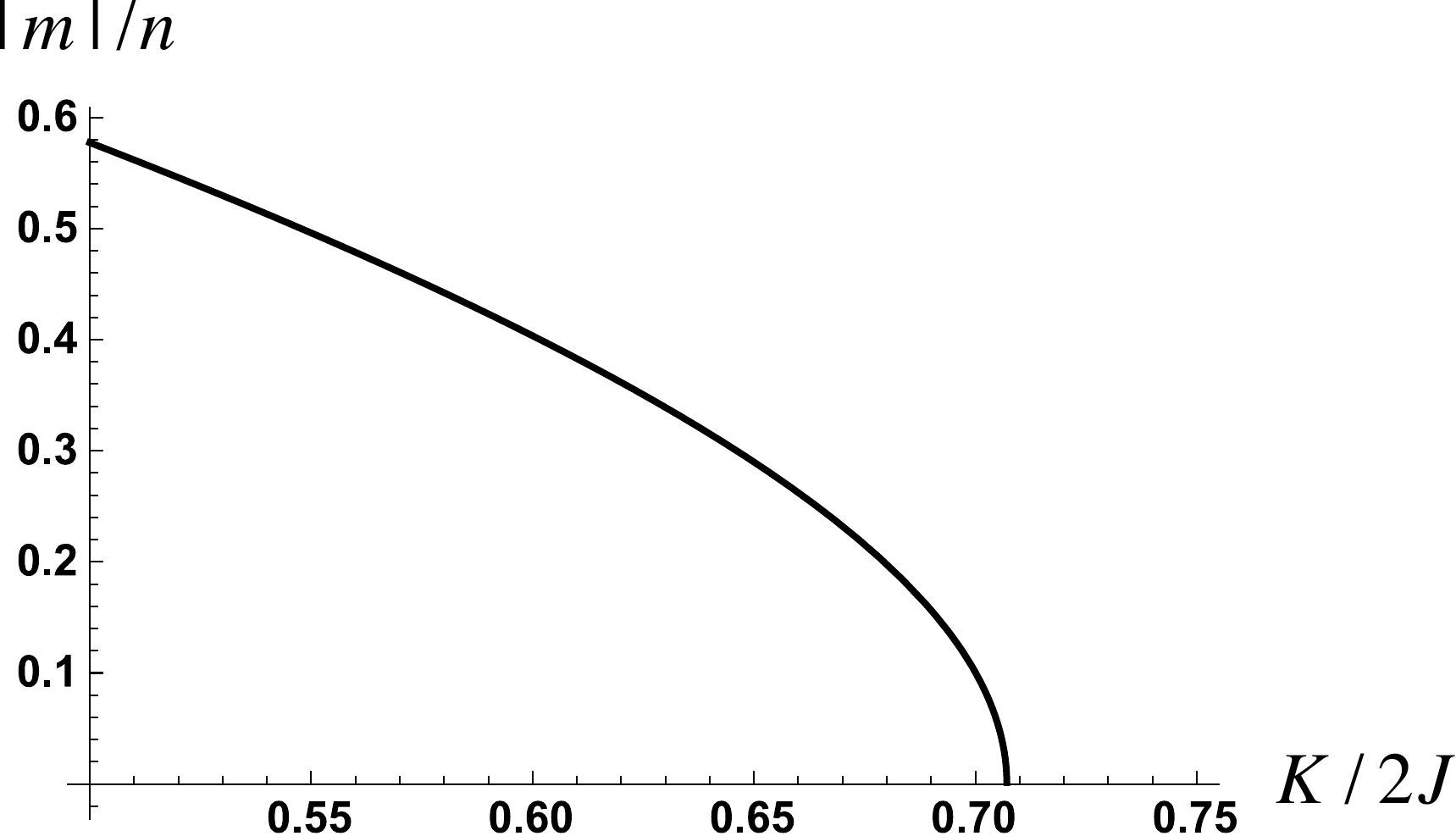}
  \caption{The absolute value of density difference at
  $\phi=\frac{\pi}{2}$ in the biased ladder phase. 
  At the boundary between biased ladder ($m\ne 0$) and Meissner phases
  ($m=0$), the density difference disappears.
  On the other hand, at the boundary between incommensurate vortex and
  biased ladder phases,  an infinite number of degeneracy in
  density difference emerges due to the emergent symmetry as shown in Sec. \ref{sec:ferroXXZ}.
  }
  \label{fig3}
 \end{center}
\end{figure}

A peculiarity of this phase is seen in the density in each leg,
$n_1$ and $n_2$.
To see this, we define a magnetization, i.e., population imbalance
between the legs, as 
$m\equiv\langle n_{1} \rangle-\langle n_{2} \rangle$.
According to the mean-field theory~\eqref{eq:gp}, the magnetization 
can be calculated as \cite{PhysRevA.89.063617}
\beq
m=- n\cos\xi_{Q}.
\label{eq:magnetization}
\eeq 
We note that it can be proved that the
magnetization value $m$ is robust even when the quantum fluctuation up
to the bosonization level is incorporated.
This is due to the fact that
the fluctuation of the magnetization $\delta m$ is given as
$\delta{m}\sim\langle{\nabla\varphi}\rangle=0$.
Thus, the mean-field result is applicable. 
As shown in Fig.~\ref{fig3}, $m$ takes a nonzero value
as far as the stability condition of the phase is met,
which means the spontaneous imbalance of the populations between the
legs occurs.
This phase corresponds to the biased ladder
phase introduced in Ref.~\cite{PhysRevA.89.063617}, which has been first
demonstrated within the GP mean-field theory. 
What is addressed here is that even in the presence of the quantum
fluctuation the $Z_2$ symmetry breaking between the populations in
the doubly-fold lowest energy in the double-well band  is maintained, and
the biased ladder phase is thus stable at the full quantum level.
This would be reasonable once one recalls the fact that even though
a continuous $U(1)$ symmetry cannot be broken in quantum one-dimensional
systems, a spontaneous breaking of a discrete symmetry is possible.

\subsubsection{Analogy with ferromagnetic XXZ model}
\label{sec:ferroXXZ}
Let us now examine the nature in the phases and transitions
between them, from the viewpoint of symmetry.
To this end, we focus on the low-energy
Hamiltonian~\eqref{eq:hamiltonian-dw}, which is invariant under the 
continuous transformations, $\beta_{\pm}\to e^{i\theta_{\pm}}\beta_{\pm}$,
and discrete transformation, $\beta_{\pm}\to\beta_{\mp}$.
This implies that symmetry of the low-energy
Hamiltonian~\eqref{eq:hamiltonian-dw} is
$U(1)_{+}\times U(1)_{-}\times Z_2$,
where the subscript $\pm$ of $U(1)$ represents the corresponding
symmetry in $\beta_{\pm}$.
We also note that the above symmetry can also be represented 
as $U(1)_V\times U(1)_A \times Z_2$ where
$U(1)_V$ and $U(1)_A$ represent the vector $U(1)$ symmetry, 
$\beta_{\pm}\to e^{i\theta}\beta_{\pm}$, and the axial $U(1)$ symmetry,
$\beta_{\pm}\to e^{\pm i\theta}\beta_{\pm}$, respectively~\footnote{The terms of  the vector and axial $U(1)$ symmetries are employed due to the analogy to chiral symmetries used in elementary particle physics.}.

In the vortex phase, 
the low-energy effective properties are captured by
the two independent TLLs, in which the two independent $U(1)$ symmetries
are hold; namely no symmetry breaking occurs in this phase.
In the biased ladder phase, on the other hand, 
the low-energy effective properties are described by the single TLL
reflecting the acoustic phonon excitation around one of the two
minimum-energy states in the double-well band. 
The $Z_2$ symmetry turns out to be broken, since the minima of the band
are degenerate and one of the minima is spontaneously chosen.

It is important to clarify the transition point between the vortex
and biased ladder phases, where we need a special attention.
First, we address that at this transition point, symmetry of
the low-energy Hamiltonian is enlarged.
To see this, it is convenient to introduce a two-component spinor,
\beq
\vec{\beta}=
\begin{pmatrix}
\beta_+\\
\beta_-
\end{pmatrix}.
\eeq
In this spinor representation, we can define $U(1)_V\times SU(2)$
transformations as
$\vec{\beta}\to e^{i\alpha_1}e^{i\alpha_2 \sigma_z}e^{i\alpha_3\sigma_y}
e^{i\alpha_4\sigma_z}\vec{\beta}$,
where  $\sigma_i$ $(i=x,y,z)$ is the Pauli matrix, and
$\alpha_{i}$ ($i=1,\cdots,4$) represents an angle of
$U(1)_V\times SU(2)$.
Then,
the low-energy Hamiltonian~\eqref{eq:hamiltonian-dw} 
is shown to be invariant under the above transformations
at the transition point where the term proportional to
$(n_+-n_-)^2$ disappears, i.e.,
the Hamiltonian has $U(1)_V\times SU(2)$ symmetry.
This implies that we can define the following conserved charges:
\beq
&&S_+=\int dx \beta^{\dagger}_+\beta_-, \label{eq:S+}\\
&&S_-=\int dx \beta^{\dagger}_-\beta_+, \label{eq:S-}\\
&&S_z=\int dx [n_+-n_-], \label{eq:Sz}\\
&&N_V=\int dx [n_++n_-]. \label{eq:Nv}
\eeq
Here, Eqs.~\eqref{eq:S+}-\eqref{eq:Sz} constitute $SU(2)$ charges
and Eq.~\eqref{eq:Nv} originates from $U(1)_V$ symmetry.
In addition, the Hamiltonian~\eqref{eq:hamiltonian-dw} 
at the transition point is identical
to the two-component bosonic Yang-Gaudin model 
where the Bethe ansatz solution is
available~\cite{li2003exact,PhysRevLett.95.150402,batchelor2006collective}.
So far the followings on two-component bosonic Yang-Gaudin model are
known: As usual, there is a TLL associated with $U(1)_V$ symmetry. 
In addition, the $SU(2)$ symmetry corresponding to Eqs. \eqref{eq:S+}-\eqref{eq:Sz}
is spontaneously broken in the ground state,
which is essentially corresponds to the physics of the Heisenberg
ferromagnet~\cite{li2003exact,PhysRevLett.95.150402,batchelor2006collective}. 
% The spontaneous breaking of the continuous symmetry in 
% this specific situation is allowed, since
% an emergence of linear NG modes in one dimension is forbidden by
% the general theorem~\cite{coleman1973there,pitaevskii1991uncertainty},
% while an emergence of quadratic NG modes is in principle possible 
% because of the lack of the precondition of the theorem.
Due to the spontaneous breaking of the $SU(2)$ symmetry,
we expect that there is a NG mode whose dispersion is quadratic in $k$ and
there exist an infinite number of the degenerate ground
states and one of them is selected spontaneously.
This is significant difference from the biased ladder phase where 
there only exists double degeneracy.

The scenario discussed above reminds us of the
similarity to the 
ferromagnetic $XXZ$ model.
When we bosonize the $XXZ$ model, the low-energy effective theory is
described as the sine-Gordon model. 
When the $XY$ in-plane anisotropy is strong, the theory is renormalized
to the TLL which corresponds to the so-called $XY$ phase, and upon
approaching the isotropic point the velocity and Luttinger parameter of
the $XY$ phase, respectively, go to zero and infinity, which implies
the quadratic dispersion of an excitation~\cite{giamarchi2003quantum}.
Furthermore, beyond the isotropic point, i.e. Ising anisotropy, the
$XXZ$ model undergoes the so-called ferromagnetic Ising phase, where
 the excitations are massive, and the $Z_2$
spin-inverse symmetry is spontaneously broken.

In our case, the anti-symmetric sector in the vortex
phase~\eqref{eq:bosonized-vx} exhibits the same effective theory as the
$XXZ$ model, and the velocity~\eqref{eq:va} and
Luttinger parameter~\eqref{eq:Ka} show the same behavior as those in the
isotropic limit of the ferromagnetic $XXZ$ model. 
Going beyond the Heisenberg point, we encounter the biased ladder phase
described by a single TLL, which corresponds to the ferromagnetic Ising
phase in the ferromagnetic $XXZ$ model. 
It is interpreted that for the biased ladder phase the anti-symmetric
sector goes away into the high enegy regime, which
would correspond to the massive excitation in the ferromagnetic Ising
phase. 
Namely such a massive excitation should describe the change of the
populations on the band bottoms and is regarded as the high energy one in
our approach, which is excluded in Eq.~\eqref{eq:Heff-BLP}.
In Sec.~\ref{sec:bogoliubov}, it is shown that 
such a massive excitation
may be incorporated with the Bogoliubov theory.

\subsubsection{Bogoliubov spectrum in biased ladder phase}
\label{sec:bogoliubov}
To see some insight into the  biased ladder phase,
let us here review the Bogoliubov theory given by~\cite{PhysRevA.89.063617}.
Since the Bogoliubov theory is based on the expansion from
the GP solution, it must underestimate fluctuation effects.
However, this does not mean that all the results by means of the
Bogoliubov theory are incorrect as pointed out in Sec. \ref{sec:single-minimum}.
At least, an excitation spectral feature in the low-energy
limit for weakly interacting one-dimensional bosons is expected to reproduce the correct behavior.
Indeed, it is known
that linear excitation feature occurring in the Lieb-Liniger model \cite{PhysRev.130.1605} 
and quadratic excitation feature occurring in the two-component Yang-Gaudin model \cite{PhysRevLett.95.150402}
can be captured by
the Bogoliubov theory.
% Based on this special attention, we implement the Bogoliubov theory.

\begin{figure}[h]
 \begin{center}
  \includegraphics[width=1\linewidth]{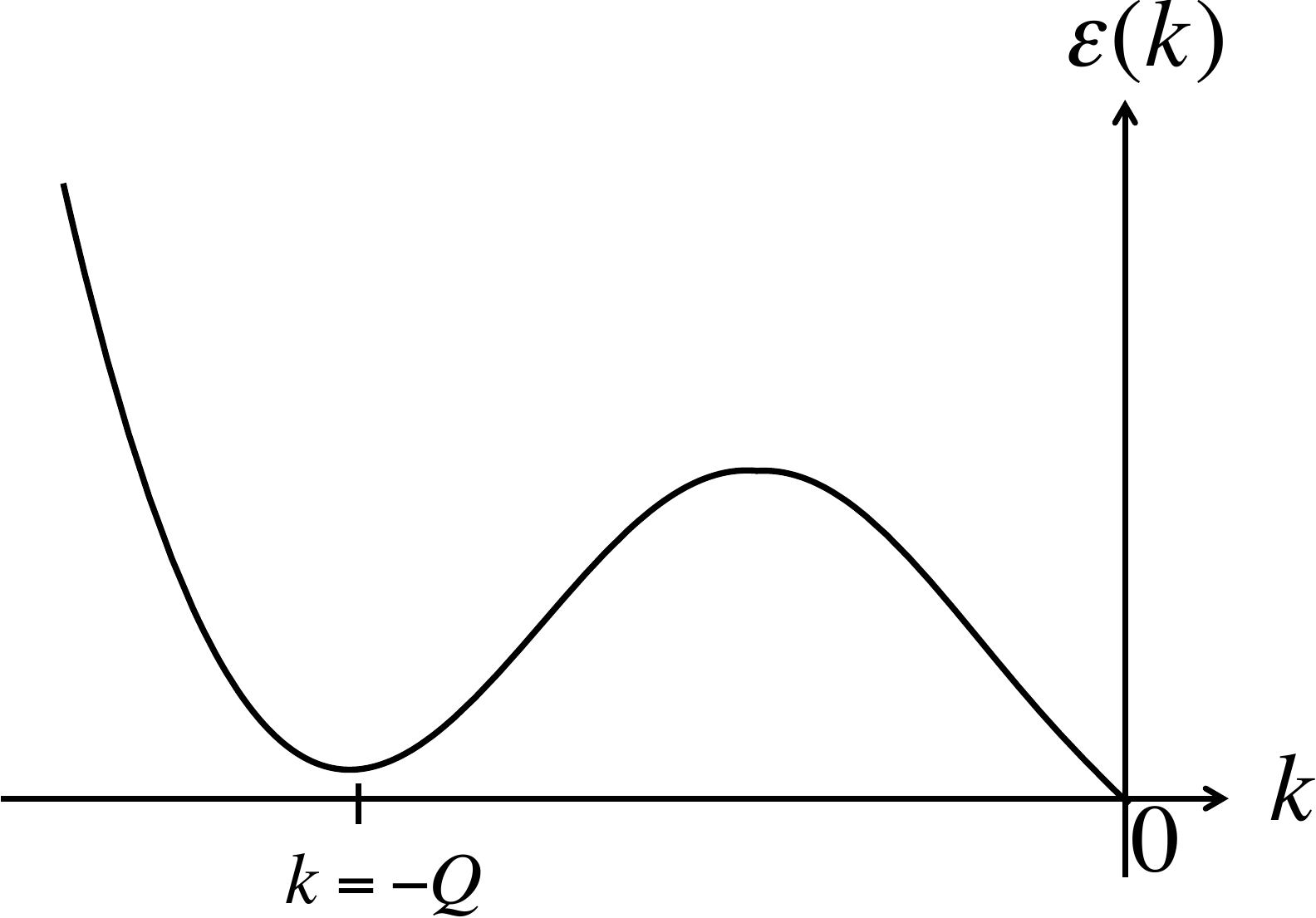}
  \caption{Schematic behavior of the Bogoliubov spectrum predicted by
  Eq.~\eqref{eq:Bogoliubov}. 
  The excitation is linear in small $k$, which should be
  interpreted as the TLL spectrum.
  On the other hand,  a local minimum exists in the vicinity of $k=-Q$
  originating from the $Z_2$ symmetry breaking.}
  \label{fig4}
 \end{center}
\end{figure}

As usual, by applying
$\beta_k=\sqrt{N_0}\delta_{k,Q}+\bar{\beta}_k$, where $\bar{\beta}_k$
is the fluctuation field and $N_0$ is the number of the particles
in the condensate, to Eq.~\eqref{eq:original-h},
we obtain the Bogoliubov Hamiltonian to be correct up to
second order in the fluctuation field 
\beq
H_{\mathrm{Bog}}&\sim&\sum_{k>0}\vec{\bar{\beta}}^{\dagger}
M\vec{\bar{\beta}}\nonumber\\
&=&
(\bar{\beta}^{\dagger}_{Q+k},\bar{\beta}_{Q-k})
\begin{pmatrix}
\zeta(k) & \eta(k) \\
\eta(k) & \zeta(-k)
\end{pmatrix}
\begin{pmatrix}
\bar{\beta}_{Q+k}\\
\bar{\beta}^{\dagger}_{Q-k}
\end{pmatrix},\nonumber\\
\label{eq:Bog-h}
\eeq
where the matrix elements are defined as 
\beq
&&\zeta(k)=E_{-}(Q+k)-E_{-}(Q)+2Un
\Big[\sin^2\frac{\xi_{Q}}{2}\sin^2\frac{\xi_{Q+k}}{2}
\nonumber\\ 
&& \ \ \ +\cos^2\frac{\xi_{Q}}{2}\cos^2\frac{\xi_{Q+k}}{2}
\Big]
-Un\Big[\sin^4\frac{\xi_{Q}}{2}
+\cos^4\frac{\xi_{Q}}{2}
\Big],\\
&&\eta(k)=Un\Big[\sin^2\frac{\xi_{Q}}{2}\sin\frac{\xi_{Q+k}}{2}
\sin\frac{\xi_{Q-k}}{2}\nonumber\\
&& \ \ \ \ \ \ \ +\cos^2\frac{\xi_{Q}}{2}\cos\frac{\xi_{Q+k}}{2}
\cos\frac{\xi_{Q-k}}{2}
\Big].
\eeq
Here, we did an approximation $N\approx N_0$.
To diagonalize the above Hamiltonian~\eqref{eq:Bog-h}, a simple scheme
is to consider the following eigenvalue
problem~\cite{blaizot1986quantum,kawaguchi2012spinor}: 
\beq
\begin{pmatrix}
1 & 0 \\
0 & -1
\end{pmatrix}
\begin{pmatrix}
\zeta(k) & \eta(k) \\
\eta(k) & \zeta(-k)
\end{pmatrix}
\begin{pmatrix}
u\\
v
\end{pmatrix}
=\epsilon
\begin{pmatrix}
u\\
v
\end{pmatrix}.
\eeq
Then, the Bogoliubov spectra $\pm\epsilon$ are obtained as eigenvalues
of the above matrix equation.
The Bogoliubov spectrum is shown to be~\cite{PhysRevA.89.063617}
\beq
\epsilon(k)=\pm\left(\frac{\zeta(k)-\zeta(-k)}{2}\right)
+\sqrt{\frac{(\zeta(k)+\zeta(-k))^2}{4}-\eta^2(k)}.\nonumber\\
\label{eq:Bogoliubov}
\eeq
A behavior of the Bogoliubov spectrum is shown in Fig.~\ref{fig4}.
When looking at $k\to0$, we have the linear spectrum, which
can be regarded as the TLL.
In addition, there is a local minimum around $k=-Q$. This roton-like
behavior would be interpreted as the massive excitation originating from
the $Z_2$ symmetry breaking.
Indeed the similar situation recently realized in
Ref.~\cite{PhysRevLett.114.055301} also occurs in a BEC in a shaken
optical lattice, in which $Z_2$ symmetry is broken.
The energy gap around $k\approx -Q$, $\epsilon(k\approx-Q)$, is found to go
closed as approaching the point where the symmetry is enlarged to $SU(2)$. 
This scenario may remind one of the ferromagnetic transition in the
$XXZ$ model when approaching the Heisenberg point from the Ising
anisotropic side, the Ising gap will collapse and the dispersion turns
to quadratic in $k$. 

\subsection{Commensurate Q case ($Q=\pi/2$)}\label{sec:Q=pi/2}
Let us next consider a commensurability effect, i.e. $Q=\pi/2$
($\phi=\pi$) case where we require considerable attention.
In this case, the double-well band structure is always 
maintained and its minima are located at $k=\pm\pi/2$
regardless of the ratio $K/J$.
A peculiarity is an emergence of the umklapp scattering process between
two energy minima in the band,
which has been overlooked so far.
To consider this, we turn back to the mean-field 
ansatz~\eqref{eq:gp}.
The form of this ansatz includes the couplings of the far-separated
states $k=\pm Q$, and thus automatically allows us to take into account
the umklapp process involving the large momentum transfer. 
% This ansatz is able to include the umklapp process explicitly, since it
% contains both $\beta_{\pm Q}^{\dagger}$ simultaneously, which allows us
% to consider the large momentum transfer.
Based on the mean-field ansatz, the ground-state energy is
straightforwardly calculated as 
\beq
&&\frac{E_0(\gamma,\theta_{\pm})}{N}=E_-(Q)+
\frac{Un}{4}\Big[-\sin^2\xi_{Q}+2\nonumber\\
&&+\left\{\left(\frac{3}{2}+\frac{\cos(2\theta_+-2\theta_-)}{2}\right)
\sin^2\xi_{Q}-1\right\}
\sin^22\gamma
\Big].
\label{eq:gs-energy-with-ukp}
\eeq
Note that it differs from Eq.~\eqref{eq:mf-energy} due to the presence
of the umklap scattering.
As can be clear from the above expression, 
the energy must be minimized when the relative phase satisfies
$\theta_+-\theta_-=\pm\frac{\pi}{2}$.
We notice that this is different from the case of a BEC in a double-well
potential, where the relative phase is zero in the ground
state~\cite{pethick2002bose}. 
This difference originates from the fact that
the relative phase dependence
is caused by a hopping (kinetic) term such as
$-J\cos(\theta_{+}-\theta_-)$
in a BEC on a double-well potential while that in our model originates from the
interaction term in our model.

By substituting this relative phase $\theta_{+}-\theta_{-}=\pi/2$ or
$-\pi/2$ into
Eq.~\eqref{eq:gs-energy-with-ukp},
the ground-state energy is going to be
\beq
\frac{E_0(\gamma)}{N}=E_-(Q)+
\frac{Un}{4}\Big[
\left(\sin^2\xi_{Q}-1\right)
\sin^22\gamma\nonumber\\
 -\sin^2\xi_{Q}+2\Big].
\eeq
Since $\sin^2\xi_{Q}\le1$,
the ground state can be uniquely determined by $\gamma=\pi/4$,
which is independent of $K/J$.
This ground state leads to the balanced density
$\langle n_+\rangle=\langle n_-\rangle$.
Thus, the biased ladder phase is washed out in the presence of the
umklapp process.

Let us next consider quantum fluctuations from the mean-field
solution.  
In this case, we need to retain the following process in the effective
Hamiltonian~\eqref{eq:hamiltonian-dw}: 
\beq
H_{\mathrm{umklapp}}=U\sin^2\xi_Q\int dx [
\beta^{\dagger}_+\beta^{\dagger}_+\beta_-\beta_-
+h.c.].
\label{eq:umklapp}
\eeq
Note that due to this term~\eqref{eq:umklapp}, symmetry of the
Hamiltonian is lowered from $U(1)\times U(1)\times Z_2$ to
$U(1)_V\times Z_2$.
Thus, the axial $U(1)$ symmetry 
($\beta_{\pm}\to e^{\pm i\theta}\beta_{\pm}$) disappears
from the low-energy Hamiltonian, and
the continuous symmetry remaining turns out to be only the vector $U(1)$
symmetry ($\beta_{\pm}\to e^{i\theta}\beta_{\pm}$).
On the other hand, the $Z_2$ symmetry ($\beta_{\pm}\to \beta_{\mp}$)
remains in the presence of the umklapp term~\eqref{eq:umklapp}.

Let us next perform bosonization as follows:
\beq
&&H=\sum_{\mu=s,a}\frac{v_{\mu}}{2\pi}\int dx
\Big[K_{\mu}(\nabla\theta_{\mu})^2+\frac{(\nabla\varphi_{\mu})^2}{K_{\mu}}
\Big]\nonumber\\
&&-\frac{g_1}{(2\pi\alpha)^2}\int dx \cos(\sqrt{8}\varphi_a) -
\frac{g_2}{(2\pi\alpha)^2}\int dx \cos(\sqrt{8}\theta_a),\nonumber\\
\label{eq:Heff-Q=pi/2}
\eeq
where $g_1=\sin^2\xi_Q/2$, $g_2=\sin^2\xi_Q/4$,
and 
\beq
&&v_s=\sqrt{\frac{nU}{M^*}},\\
&&v_a=\sqrt{\frac{nU(1-\sin^2\xi_Q)}{M^*}},\\
&&K_s=\sqrt{\frac{n}{4M^*U}},\\
&&K_a=\sqrt{\frac{n}{4M^*U(1-\sin^2\xi_Q)}}.
\eeq
An essential difference from the incommensurate $\phi$ case
is the presence of $\cos\sqrt{8}\theta_a$ which comes from
$H_{\mathrm{umklapp}}$.
Furthermore we move on to the renormalization group analysis to see the
low-energy properties of the system.
The parameters in Eq.\eqref{eq:Heff-Q=pi/2} are found to obey the
following renormalization group equations as (See Appendix)
\beq
\frac{d(g_1/v_a)}{dl}=2(1-K_a)g_1/v_a,\label{eq:RG1}\\
\frac{d(g_2/v_a)}{dl}=2\left(1-\frac{1}{K_a}\right)g_2/v_a
\label{eq:rg-commensurate}.
\eeq
Since $K_a\gg1$ in the weakly interacting case assumed here, it is found
from Eqs.~\eqref{eq:RG1} and~\eqref{eq:rg-commensurate} that $g_1/v_a$
and $g_2/v_a$ are rapidly renormalized, respectively, to being zero and
divergent as $l$ increases. 
Namely, $\cos\sqrt{8}\theta_a$ in Eq.~\eqref{eq:Heff-Q=pi/2} is highly
relevantly retained in the effective Hamiltonian in the low-energy limit
while $\cos\sqrt{8}\phi_a$ irrelevantly goes away as in the case of
$Q\ne\pi/2$: The renormalized effective theory is
\beq
H=\sum_{\mu=s,a}\frac{v_{\mu}}{2\pi}\int dx
\Big[K_{\mu}(\nabla\theta_{\mu})^2+\frac{(\nabla\varphi_{\mu})^2}{K_{\mu}}
\Big]\nonumber\\
-\frac{g_2}{(2\pi\alpha)^2}\int dx \cos(\sqrt{8}\theta_a).
\label{eq:Heff-Q=pi/2--2}
\eeq
Thus, the anti-symmetric sector becomes gapful due
to the fixed relative phase $\langle{\cos(\sqrt{8}\theta_a)}\rangle=1$,
which is consistent with the mean-field scenario, deduced by
Eq.~\eqref{eq:gs-energy-with-ukp}, with the umklapp scattering,

Let us also look at the currents in this phase.
By means of the bosonization formula, 
the averaged currents are calculated as
\cite{tokuno}
\beq
\langle j_c(x)\rangle&\sim& 2nJ\sin{\xi_Q}\,
(-1)^x,\\
\langle j^{\perp}(x)\rangle&\sim& nK\cos{\xi_{Q}}\,(-1)^x.
% \langle j_c(x)\rangle&\sim& 4nJ\sin\left(\frac{\xi_Q}{2}\right)
% \cos\left(\frac{\xi_Q}{2}\right)(-1)^x,\\
% \langle j^{\perp}(x)\rangle&\sim& -nK\Big[
% \sin^2\left(\frac{\xi_Q}{2}\right)-\cos^2\left(\frac{\xi_Q}{2}\right)
% \Big](-1)^x.
\eeq
Thus, the current pattern shows up due to
the umklapp effects.
In a similar manner, we can evaluate the current correlations,
which turn out to be zero (or exponentially decay in $x$).
This phase is called (commensurate) vortex or chiral superfluid
phase.

\begin{figure}[t]
 \begin{center}
  \includegraphics[width=1\linewidth]{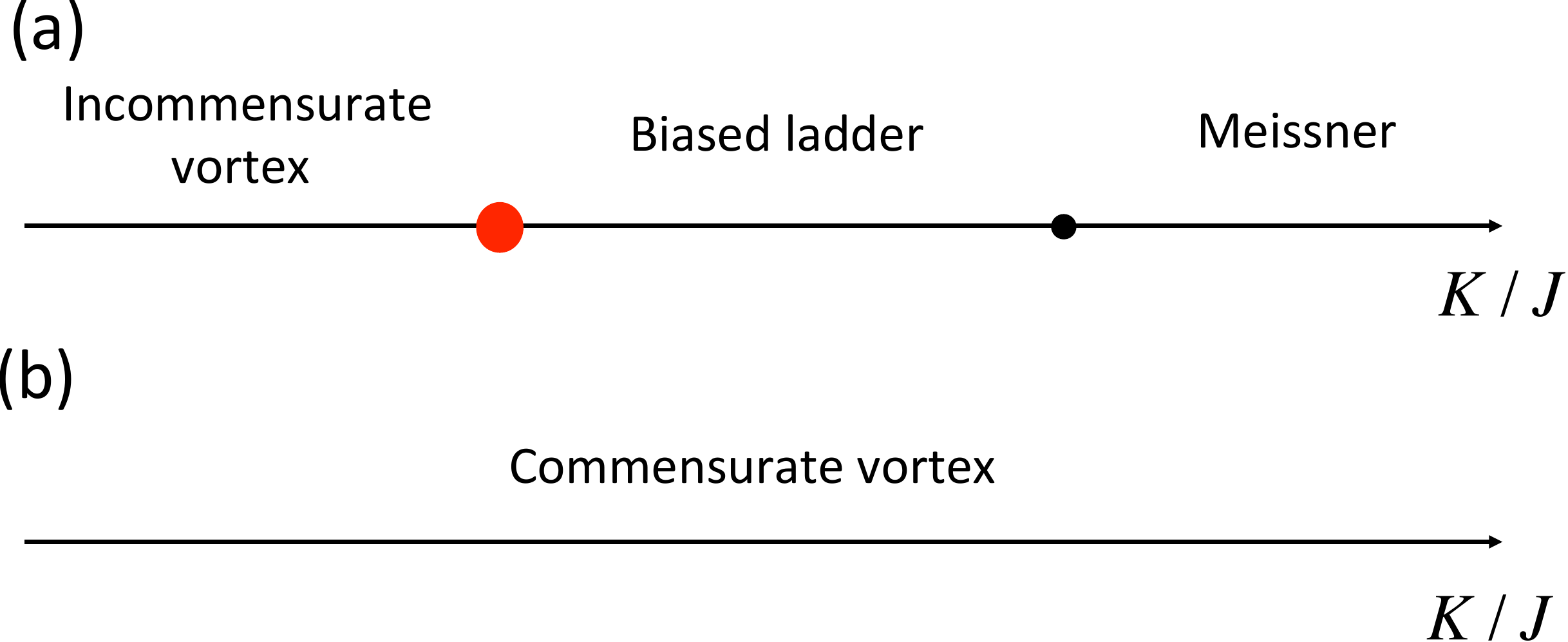}
  \caption{Schematic phase diagram (a) for $Q\ne\pi/2$ and (b) for
  $Q=\pi/2$.
  An emergent $SU(2)$ symmetry shows up at the boundary between
  incommensurate vortex and biased ladder phases, which is marked with the red circle in (a).} 
  \label{fig5}
 \end{center}
\end{figure}

The possible phase diagram in the weak coupling regime
$J,K\gg U$ is summarized in Fig.~\ref{fig5}.

\subsubsection{Transition between commensurate and incommensulate fluxes}
Now we discuss properties in the vicinity of the commensurate flux 
$\phi=\pi$.
To this end, we first estimate the gap coming from 
the coupling $\cos\sqrt{8}\theta_a$ in Eq.~\eqref{eq:Heff-Q=pi/2--2}
based on the renormalization group
equation~\cite{giamarchi2003quantum}.
As mentioned above, this coupling $g_2$ is highly relevant in $K_a\gg1$
and is therefore expected to flow to the strong coupling regime
($g_2 \rightarrow\infty$).
Thus, we introduce a typical length scale $l^*$ estimated from
Eq.~\eqref{eq:rg-commensurate} as 
\beq
e^{l^*}\sim\left(\frac{v_2}{g_2}\right)^{\frac{1}{2-\frac{2}{K_a}}},
\eeq
and stop the renormalization group flow at the length
scale $l^{*}$.
On the other hand, if we are in the strong coupling,
we may approximate the cosine term as
$g_2\cos(\sqrt{8}\theta_a)\approx g_2 [1 - 4\theta_a^2(x)]$.
Therefore, applying the expansion to the renormalized effective
theory~\eqref{eq:Heff-Q=pi/2--2}, the Hamiltonian is easily
diagonalized, and the consequently estimated gap $\Delta(l^*)$ for the
renormalized coupling constant $g_2(l^{*})$ at a termination of
the renormalization group flow is found to be
$\Delta(l^*)\sim\sqrt{\frac{g_2(l^*)v_a}{K_a}}\sim\sqrt{\frac{v_a}{K_a}}$.
Taking into account that the gap is renormalized as
$\Delta(l)=e^l\Delta$, we obtain
\beq
\Delta\sim g^{\frac{1}{2-\frac{2}{K_a}}}_2\sqrt{\frac{v_a}{K_a}}.
\eeq

Based on this estimation, 
we next consider the situation in which $Q$ is slightly deviated from
a commensurate point $Q=\pi/2$, and write $Q$ as
$Q=\frac{\pi}{2}+\frac{\delta }{4}$, where $\delta$ is assumed to be
small. 
Then, the umklapp term~\eqref{eq:umklapp} is
bosonized as 
\beq
H_{\mathrm{umklapp}} \sim - \frac{g_2}{(2\pi\alpha)^2}\int dx
\cos(\sqrt{8}\theta_a+\delta x),
\eeq
which is namely the oscillating term as a function of $x$.
We may safely make the replacement of $\delta \rightarrow 0$
if $\delta$ is irrelevant.
However, if $\delta$ is relevant, the umklapp term experiences a strong
oscillation, and cancels out in the renormalized Hamiltonian.
In general, the transition from the irrelevant and
relevant $\delta$ or vice versa is known to occur around
$\Delta\sim \delta$, and called a commensurate-incommensurate
transition~\cite{giamarchi2003quantum}.
In our cases, the transition between the chiral superfluid and vortex
phases corresponds to such a commensurate-incommensurate
transition~\cite{giamarchi2003quantum},
because once $\delta$ becomes relevant, the effective theory reduces to
that of the two independent TLLs, which is identical to that of the
vortex phase. 

On the other hand, the transition between the commensurate vortex
and biased ladder phases may not be captured by this scenario of the
incommensurate-commensurate transition because the effective theory in
the biased ladder phase is not the two independent TLLs.

It is also interesting to consider the transition between the commensurate vortex
and Meissner phases.
However, it is difficult to describe such a transition by means of our approach.
Thus, it would be worthwhile examining the nature of these transitions in
a numerical simulation.

\subsection{Other commensurability effect}~\label{sec:otherC}
So far, we have discussed only the $\phi=\pi$ case as a commensurate case.
The point on the $\phi=\pi$ case is that the biased ladder phase is
suppressed by the presence of the umklapp scattering.
Here we consider the other commensurate case　
in order to see roles of general commensurability, and we will
see that the $\phi=\pi$ commensurability is special. 

As heretofore, let us first consider the mean-field approximation for the Hamiltonian~\eqref{eq:original-h}.
Then, it turns out that as far as such a Hamiltonian is concerned, 
we always have the energy expression~\eqref{eq:mf-energy} regardless of commensurability of $Q$.
Thus, the mean-field phase diagram is identical to the incommensurate $Q$ case.

Let us next consider the bosonization to see quantum fluctuation effects.
For the incommensurate vortex phase at the mean-field level, it is expected that
the higher order perturbation theory generates cosine terms in
$\theta_a$ at a commensurate $Q$ as shown in Ref.~\cite{PhysRevB.64.144515}.
Since many of such cosine terms are relevant for $K_a\gg1$,
we find that the incommensurate vortex phase is replaced by 
the commensurate vortex phase by quantum fluctuation effects
if the coupling of the cosine term is larger than the temperature~\cite{giamarchi2003quantum}.

On the other hand, for the biased ladder phase at the mean-field level,
we find that a cosine term to fix $\theta_a$ does not show up by the bosonization since
the mean density in one of the wells is equal to zero.
Namely, the biased ladder phase at such a commensurate $Q$ is robust.

\section{Summary and Perspective}
\label{sec:summary}
We have examined the two-leg Bose-Hubbard ladder model subject to a
magnetic field flux. 
We have particularly revealed the structure of the phase diagram in a
weak-coupling regime by using a couple of the effective theory methods.
What we stress is that we have also found the so-called biased-ladder
phase, first predicted by the GP mean-field approach, to be robust against 
quantum fluctuations.
It has also been shown that the transition between the biased-ladder and vortex phases has a
similarity as that of the ferromagnetic $XXZ$ model where the emergent $SU(2)$ symmetry 
comes out at the transition between the biased ladder and vortex phases.
In the case of the ladder system subject to the magnetic flux, 
commensurability works to phase degrees of freedom, which produces
a kind of umklapp processes. 
By incorporating such an umklapp process at the mean-field
level, we have shown that the biased-ladder state tends to be
destabilized by the umklapp process, and turns out to be forbidden for
the case of $\phi=\pi$.

\subsection{Transition between Meissner and biased ladder phases}
As seen in Sec.~\ref{sec:formulation}, the dispersion becomes
quartic at the critical point between the single- and double-minimum
band structures.
In the absence of an interaction, one can naively expect that
all the bosons condense at the lowest energy, and just forms a BEC which
is the same as the case of the quadratic dispersion.

A question is what happens in the presence of an interaction.
Here we briefly discuss a possible scenario.

We first point out that the similar situation can be also considered for
one-dimensional two-component bosons with spin-orbit couplings, where
the bare 
single-particle dispersion becomes quartic at a certain value of
the spin-orbit coupling and biased chemical potential between the two
species.
By employing the hydrodynamic approach and Gaussian approximation,
it is shown that the low-energy effective theory undergoes
non-TLL~\cite{PhysRevA.90.011602}. 
Then the excitation is still gapless, but is no longer identical to
an acoustic phonon: the quadratic-dispersion mode. 
It would rather be that of the Heisenberg ferromagnet and two-component
bosonic Yang-Gaudin model where the spontaneous symmetry breaking and NG
mode show up. 
Interestingly, however, the off-diagonal density matrix is shown to
decay exponentially as
\beq
\langle b^{\dagger}_{x,p}b_{0,p'} \rangle\sim
ne^{-|x|/\xi_c},
\eeq
where $\xi_c$ is the correlation length given by
$\xi_c=\sqrt{2\rho_0/(mg\lambda^2)}$ with a mean-density $\rho_0$,
atomic mass $m$, density-density interaction $g$ and spin-orbit coupling
$\lambda$.
% $\xi_c=\sqrt{\frac{8n\alpha}{U}}$ with the coefficient of 
% the single particle dispersion $\alpha$.
Thus, it means that even one-dimensional superfluidity is destroyed.

From the above example, we can expect the same physics in our model.
Namely the system might form such a non-TLL when the system transits
from the Meissner to the biased ladder state.
However, then the Meissner current would be predicted to be still
protected because its presence is guaranteed by the two band
structure from the ladder geometry and flux~\eqref{eq:band}.~\cite{tokuno}

In addition to the physical properties, this problem on interacting
bosons for quartic bare dispersion would have another interesting
aspect.
In general, it is expected that an existence of gapless modes supports LRO
or quasi-LRO while our model is an exceptional case on this statement.
Thus, the profound understanding on the role of gapless modes in one dimension remains an
open question.

\subsection{Stronger $U$ effect}
In the paper, we have analysed the system under the condition
$K,J\gg U$.
A natural question to come up then is 
what happens when the system with a stronger interaction $U$ is
concerned.
In many of one-dimensional systems, both the weak- and strong-coupling
analyses are continuously connected to each
other consequently~\cite{giamarchi2003quantum},
but we would think that the strong coupling regime includes different
physics in our model.
This is because while in a weak coupling,
the low-energy properties can be well captured by assuming the
quasi-condensates at well-separated lowest energy single particle states
in the double-well band structure, such a picture is no longer applicable
in a strong coupling due to a hard-core feature analogous to
\textit{Fermi statistics}.
Namely the double-well band feature can be no longer important in
low-energy physics if we naively assume fermion-like occupation of the
particles in the band picture as a strong coupling limit, and the
different properties from those for the weak coupling should be then
found.
Indeed, the recent numerical analysis~\cite{PhysRevB.91.140406} in a strong coupling
shows that the biased-ladder phase is not found while the presence of
the Meissner and vortex phases is confirmed.

Let us make a further consideration on the physics in the intermediate
interaction.
Then it is convenient to take the generalized mean-field
ansatz~\cite{PhysRevA.89.063617},
\beq
|GS'\rangle=\frac{1}{\sqrt{N!}}\left(
e^{i\theta_+}\cos\gamma\beta_k^{\dagger}+e^{i\theta_-}\sin\gamma
\beta_{k}^{\dagger}\right)^N|0\rangle,
\eeq
where now the wave-vector $k$ pointing at the bottoms of the band
is also treated as a variational parameter.
This generalization means that a modification of the band structure by
an interaction is taken into account.
As shown in Ref.~\cite{PhysRevA.89.063617}, the variational approach
shows that the optimized value of $k$ decreases with $U$, and approaches
zero at a certain $U_c$. 
It is not clear whether this mean-field ansatz correctly captures the
physics in the regime $U\sim U_c$, but we can naively guess, at least,
from this discussion, that the interaction works so as to collapse the
double-well band structure.

Combining the mean-field and numerical result, the following scenario can
be deduced: 
The biased-ladder state for the weak-interaction goes unstable as $U$
increases; it eventually transits at $U=U_c$ to Meissner state, and 
such a Meissner state continues to that of the strong coupling regime
which is found in Ref.~\cite{PhysRevB.91.140406}.
To test this scenario, or to precisely estimate the critical $U_c$, an
unbiased numerical simulation would be necessary.

\textit{Note added:}
Recently,
we noticed a paper \cite{2015arXiv150406564G}, which found
the biased ladder phase in a regime $J\simeq K\simeq U$ by means of the
density matrix renormalization group.

\begin{acknowledgements}
 S.U. is supported by the Swiss National Science Foundation under
 Division II. 
\end{acknowledgements}

\appendix

\begin{widetext}
\section{Derivation of renormalization group equation}
In this Appendix, we derive renormalization group equations
for the following Hamiltonian:
\beq
&&H=H_{0}+\frac{\lambda_1}{(2\pi\alpha)^2}\int dx\cos(\sqrt{8}\varphi)
+\frac{\lambda_2}{(2\pi\alpha)^2}\int dx\cos(\sqrt{8}\theta),
\label{eq:fh-rg}\\
&&H_0=\frac{v}{2\pi}\int dx\left[\frac{1}{K}(\nabla\varphi)^2
  +K(\nabla\theta)^2\right],
\label{eq:tlh-rg}
\eeq
where $\lambda_1$ and $\lambda_2$ are the couplings of the cosine terms, and
the cutoff parameter $\alpha$ turns out to play an important role in
obtaining renormalization group equations.
To this end, we adopt a scheme based on correlation functions
\cite{PhysRevB.16.1217}, which
is known to be useful in many one dimensional systems
\cite{giamarchi2003quantum}.

To be specific, we consider the following correlation function:
\beq
R(r_1-r_2)=\langle e^{i\sqrt{2}\varphi(x_1,\tau_1)}e^{-i\sqrt{2}
  \varphi(r_2,\tau_2)} \rangle,
\eeq
where $r_i=(x_i,y=v\tau_i)$ ($i=1,2$), and the average is taken
with the following partition function:
\beq
Z=\int D\varphi D\theta
e^{\int d\tau dx\left[\frac{i\nabla\theta\partial_{\tau}\varphi}{\pi}-H\right] }.
\label{eq:partition-function}
\eeq
If the couplings
$\lambda_1$ and
$\lambda_2$ are absent, the Hamiltonian becomes the TLL one (Gaussian),
and therefore, we can easily evaluate the above correlation function
for $r_1-r_2\gg\alpha$ as
\beq
\langle e^{i\sqrt{2}\varphi(x_1,\tau_1)}e^{-i\sqrt{2}
  \varphi(r_2,\tau_2)} \rangle_{H_0}\sim e^{-KF_1(r_1-r_2)},
\eeq
where
$F_1(r)=\frac{1}{2}\ln\Big[\frac{x^2+(v|\tau|+\alpha)^2}{\alpha^2}\Big]$, and
$\langle \cdots \rangle_{H_0}$ means that
the average is taken with $H_0$ in Eq. \eqref{eq:partition-function}.
While in the presence of the couplings,
one cannot evaluate the partition function
exactly, at least,
one can perform a perturbative calculation by assuming that
the couplings are small.
Then, it is easily to show that the first-order terms in the couplings become
zero.
Thus, up to the second-order terms, the correlation function is given by
\beq
R(r_1-r_2)=e^{-KF_1(r_1-r_2)}\Big[1+\frac{\lambda_1^2}{2(2\pi\alpha)^4v^2}
  \sum_{\epsilon=\pm1}\int d^2r'd^2r'' e^{-KF_1(r'-r'')}
  (e^{2K[F_1(r_1-r')-F_1(r_1-r'')+F_1(r_2-r'')-F_1(r_2-r')]}-1)
  \nonumber\\
+\frac{\lambda_2^2}{2(2\pi\alpha)^4v^2}
  \sum_{\epsilon=\pm1}\int d^2r'd^2r'' e^{-K^{-1}F_1(r'-r'')}
  (e^{-2[F_2(r_1-r')-F_2(r_1-r'')+F_2(r_2-r'')-F_2(r_2-r')]}-1)
  \Big],\nonumber\\
\eeq
where $F_2(r)=-i\text{Arg }(v\tau+\alpha+ix)$,
and $d^2r=vdxd\tau$.
To go further, let us use the fact that dominant contributions
in the above integrals comes from regions where $r'$ and $r''$ are
not too distant.
Thus, by introducing the center of mass and relative coordinates,
$R=\frac{r'+r''}{2}$ and $r=r'-r''$, 
one can expand the exponential terms in the parentheses and
obtain
\beq
&&R(r_1-r_2)=e^{-KF_1(r_1-r_2)}\Big[1\nonumber\\
&&  -\frac{\lambda_1^2}{2(2\pi\alpha)^4v^2}
  \int d^2Rd^2r e^{-KF_1(r)}K^2r^2
  (F_1(r_1-R)-F_1(r_2-R))(\nabla^2_X+\nabla^2_Y)(F_1(r_1-R)-F_1(r_2-R))
  \nonumber\\
&&-\frac{\lambda_2^2}{2(2\pi\alpha)^4v^2}
  \int d^2Rd^2r e^{-K^{-1}F_1(r)}r^2
  (F_2(r_1-R)-F_2(r_2-R))(\nabla^2_X+\nabla^2_Y)(F_2(r_1-R)-F_2(r_2-R))
  \Big].\nonumber\\
\eeq
By using the so-called Cauchy relations,
\beq
\nabla_XF_1=i\nabla_YF_2,\\
\nabla_YF_1=-i\nabla_XF_2,
\eeq
and $(\nabla^2_X+\nabla^2_Y)\ln(R)=2\pi\delta(R)$,
the correlation function is given by
\beq
R(r_1-r_2)=e^{-KF_1(r_1-r_2)}\Big[1+\frac{\lambda_1^2KF_1(r_1-r_2)}{4\pi^3\alpha^4 v^2}
  \int d^2rr^2 e^{-KF_1(r)}
-\frac{\lambda_2^2F_1(r_1-r_2)}{4\pi^3\alpha^4v^2}
  \int d^2rr^2 e^{-K^{-1}F_1(r)}
  \Big].
\eeq
This expression allows one to introduce an effective exponent $K_{\text{eff}}$,
\beq
K_{\text{eff}}=K-\frac{y^2_1K^2}{2}\int \frac{dr}{\alpha}
\left(\frac{r}{\alpha}\right)^{3-4K}
+\frac{y^2_2}{2}\int \frac{dr}{\alpha}
\left(\frac{r}{\alpha}\right)^{3-\frac{4}{K}},
\eeq
where we introduced $y_i=\frac{\lambda_i}{\pi v}$ ($i=1,2$).
Considering that the effective exponent should not affect
by varying the cutoff as $\alpha'=\alpha+d\alpha$, we obtain
\beq
&&K(\alpha')=K(\alpha)-\frac{y^2_1(\alpha) K^2(\alpha)}{2}\frac{d\alpha}{\alpha}
+\frac{y^2_2(\alpha)}{2}\frac{d\alpha}{\alpha},\\
&&y^2_1(\alpha')=
y^2_1(\alpha)\left(\frac{\alpha'}{\alpha}\right)^{4-4K(\alpha)},\\
&&y^2_2(\alpha')
=y^2_2(\alpha)\left(\frac{\alpha'}{\alpha}\right)^{4-\frac{4}{K(\alpha)}}.
\eeq
By introducing the scaling parameter $l$, which satisfies $\alpha=\alpha_0e^l$
with the original cutoff $\alpha_0$, the following renormalization group
equations are obtained:
\beq
&&\frac{dK}{dl}=-\frac{K^2y^2_1}{2}+\frac{y^2_2}{2},
\label{eq:rg1}\\
&&\frac{dy_1}{dl}=2(1-K)y_1,
\label{eq:rg2}\\
&&\frac{dy_2}{dl}=2(1-1/K)y_2.
\label{eq:rg3}
\eeq

The low-energy effective theory~\eqref{eq:bosonized-vx} in the
incommensurate vortex phase corresponds to Eq.~\eqref{eq:rg-invortex}
with the constraint $\lambda_2=0$.
Thus by fixing $y_2$ to be zero in Eqs.~\eqref{eq:rg1}-\eqref{eq:rg3} and
replacing $y_1=g/v_a$, the renormalization group
equation~\eqref{eq:rg-invortex} is derived. 
On the other hand, the effective theory~\eqref{eq:Heff-Q=pi/2} in the
commensurate vortex phase is identical to Eq.~\eqref{eq:fh-rg}, and thus
the renormalized group equations, Eqs.~\eqref{eq:RG1}
and~\eqref{eq:rg-commensurate}, are exactly obtained by the replacement
$y_1=g_1/v_a$ and $y_2=g_2/v_a$. 

\end{widetext}

%\bibliographystyle{apsrev4-1}
%\bibliography{reference}
%

\end{document}